\documentclass[floatfix,aps,prd,amsmath,nofootinbib,onecolumn,11pt]{revtex4-1}
\def \be {\begin{equation}} 
\def \ee {\end{equation}} 
\def \bea {\begin{eqnarray}} 
\def \eea {\end{eqnarray}} 

\usepackage{graphicx}
\usepackage{dcolumn}
\usepackage{bm}
\usepackage{epsfig} 
\usepackage{amsfonts}
\usepackage{amsmath}
\usepackage{amssymb}
\usepackage[usenames]{color}
\usepackage[dvipsnames]{xcolor}
\usepackage{float}
\usepackage[unicode, colorlinks=true, linkcolor=linkcolor, citecolor=linkcolor, filecolor=linkcolor,urlcolor=linkcolor, pdfusetitle]{hyperref}

\hypersetup{colorlinks,citecolor=blue,linkcolor=blue,urlcolor=blue}
\hypersetup{final=true}

\begin{document}

\title{Current constraints on the minimally extended varying speed of light model through the cosmic distance duality relation}

\author{Jaiane Santos}
\email{jaianesantos@on.br}
\affiliation{Observat\'orio Nacional, 20921-400, Rio de Janeiro - RJ, Brazil}

\author{Carlos Bengaly}
\email{carlosbengaly@on.br}
\affiliation{Observat\'orio Nacional, 20921-400, Rio de Janeiro - RJ, Brazil}

\author{Rodrigo S. Gonçalves}
\email{rsg\underline{ }goncalves@ufrrj.br}
\affiliation{Departamento de Física, Universidade Federal Rural do Rio de Janeiro, 23897-000, Seropédica - RJ, Brazil}
\affiliation{Observat\'orio Nacional, 20921-400, Rio de Janeiro - RJ, Brazil}

\date{\today}

\begin{abstract}
One of the most crucial tests of the standard cosmological model consists on testing possible variations on fundamental physical constants. In frameworks such as the minimally extended varying speed of light model (meVSL), the relationship between the luminosity distance ($D_{\text{L}}$) and the angular diameter distance ($D_{\text{A}}$), namely the cosmic distance duality relation (CDDR), is expected to deviate from $\eta(z) \equiv D_{\text{L}}/D_{\text{A}}(z)(1+z)^{-2} = 1$, making it a powerful probe to check a potential variation of fundamental constants. Hence, we test the viability of the meVSL model through the CDDR by comparing $D_{\text{A}}$ measurements, provided by the transverse (2D) and anisotropic (3D) baryon acoustic oscillations (BAO) observations from different surveys, like SDSS, DES and DESI, in combination with $D_{\text{L}}$ measurements from Pantheon+ type Ia Supernova (SNe) compilation. The Gaussian Process reconstruction is employed on the SN data to match $D_{\text{A}}$ with $D_{\text{L}}$ at the same redshift. We find no deviation of the standard CDDR relation within 1-2.6$\sigma$ confidence level when considering SNe with 2D and 3D BAO samples combined together. However, when SNe and 2D BAO only are considered, the standard CDDR is only recovered at $\sim 4\sigma$ confidence level. However, such a result might be due to some recently discussed tensions between SN and BAO datasets, especially at low redshifts, in addition to possible inconsistencies between the BAO datasets individually. Therefore, our results show no significant evidence in favour of the meVSL model, once these potential systematics are taken into account.

\end{abstract}

\keywords{Cosmology: Theory -- Cosmology: Observations -- Type Ia Supernovae -- BAO}

\pacs{98.65.Dx, 98.80.Es}
\maketitle

\section{Introduction}\label{sec:intro}
The standard cosmological model (SCM) has also been known as $\Lambda$CDM, since the late 1990s \cite{Riess,Perlmutter}, where $\Lambda$ is the cosmological constant, responsible for the accelerated expansion of the Universe and CDM, refers to cold dark matter, a component that interacts only gravitationally and which plays a fundamental role in the formation of structures and dynamics of galaxies. The SCM is based on two main features, the first is the assumption of a Universe described by the Friedman-Lemaître-Robertson-Walker (FLRW) metric on large scales \cite{Wu:1998ad,Clarkson:2010uz,Maartens:2011yx,Clarkson:2012bg}, and the second is the theory of General Relativity as the theory of the gravity behind this model \cite{Clifton:2011jh,Will:2014kxa,Berti:2015itd,Coley:2019yov}. The SCM scenario provides several successful predictions, for example, recent observations of the Cosmic Microwave Background (CMB) \cite{Aghanim}, Type Ia Supernovae (SNe) luminosity distances \cite{SCOLNIC}, as well as galaxy clustering and weak lensing \cite{Alam,Asgari,Abbott,Secco}, confirm this model as the one that best describes the data.

Although it is well established, currently the SCM suffers from some unsolved problems, such as the value of the vacuum energy density \cite{Weinberg:1988cp,Padmanabhan:2002ji}, the primordial singularity problem \cite{Baumann:2018muz} and observational discrepancies between measurements of cosmological parameters, which arise from different scales observation \cite{DiValentino:2020zio,DiValentino:2020vvd}. At present, the $\sim$ 5$\sigma$ tension between the Hubble Constant, $H_{0}$, measured in the near Universe with SNe and in the early one with CMB ~\cite{Riess:2021jrx,Shah:2021onj}, is one of the most intriguing questions. Therefore, it is essential to propose and test alternative models and revisit the foundations of the SCM. Given that additional evidence of deviations from the SCM predictions could indicate new physics at play, a complete reformulation of its framework would be required, for instance scenarios with hot axions~\cite{Deramo:2018gsr},  Chameleon dark energy ~\cite{Vagnozzi:2021gws}, among others - for a complete review on possible solution we refer to~\cite{DiValentino:2021izs, Perivolaropoulos:2022rge, HuWang:2023sdf}.

A possible route to verify the consistency of the $\Lambda$CDM model is to study a possible violation of the cosmic distance duality (CDDR) \cite{ELLIS}, as it corresponds to one of the key relations in cosmology, and is closely related to Etherington's theorem. This relation connects two  measures of cosmic distance for an astronomical object: the luminosity distance $D_{\text{L}}$ and the angular diameter distance $D_{\text{A}}$, in a spacetime described by a metric theory of gravity where the number of photons is conserved as they travel along null geodesic, which reads
\begin{equation}
\label{eq:eta1z}
    \eta(z) = \frac{D_{\text{L}}}{D_{\text{A}}}(1+z)^{-2} = 1.
\end{equation}
with $z$ representing the redshift of the astronomical object.

There are several possibilities for a violation of the CDDR. Among them, we have non-conservation of the number of photons due to coupling to particles beyond the standard particle physics model, absorption of dust \cite{Csaki:2001yk,Bassett:2003zw,Lima:2011ye}, photons not traveling along null geodesics \cite{Schuller:2017dfj,Schneider:2017tuk}, as well as varying fundamental constants such as the fine-structure constant~\cite{Goncalves:2019xtc,Bora:2020sws} and the speed of light~\cite{Ellis:2007ah} -- for a broad review on this topic, we refer the interested reader to~\cite{Uzan:2024ded}. 

As a general perspective, the possible violations mentioned arise from the violation of any physical mechanisms behind the CDDR, those assumptions are basically that many geometric properties
are invariant under transposition between observer and source. Those results can be proven under the FLRW cosmology, in a Riemannian spacetime and with light
travelling on null geodesics~\cite{ELLIS}. Hence, observational tests of the CDDR constitutes a crucial probe of the standard model, as well as of fundamental physics.

In recent literature, many approaches have been designed and pursued to test the validity of the CDDR in the context of some astrophysical object. For example, Gamma-ray bursts (GRBs) were used to test the validity of the CDDR, as they can be used as standard candles and provide luminosity distances at high redshifts \cite{Holanda:2014lna}. From gravitational waves as standard sirens, the luminosity distance can be obtained without the need for a cosmic distance ladder, thus serving as a test for the CDDR \cite{Yang:2017bkv,Qi:2019spg,Arjona:2020axn}. As for the Cosmic Microwave Background (CMB), any deviation of its black body spectrum would result in a violation of the CDDR \cite{Ellis:2013cu}. Type Ia Supernovae (SNe) data have also been used as a source of luminosity distance to test the CDDR, as in \cite{Bassett:2003vu}. Notably, strongly lensed SNe can provide both the luminosity distance and angular diameter distance, thus enabling a CDDR test \cite{Renzi:2020bvl}. Another way to test the CDDR is by means of the relationship between the gas mass fraction of galaxy clusters (GCs) obtained through the Sunyaev-Zeldovich effect ($f_{\text{SZE}}$) and through the X-ray surface brightness ($f_{\text{X}}$), which is given by $f_{\text{SZE}}/f_{\text{X}} = D_{\text{L}}/(D_{\text{A}}(1+z)^{2}) \equiv \eta(z)$ \cite{Holanda:2012at}. Combinations of different cosmological probes have been used to test the CDDR, e.g. the luminosity distance and angular diameter distance measurements from SNe and GCs, respectively \cite{Holanda:2010vb,Li:2011exa,Meng:2011nt,Holanda:2011hh,Goncalves:2011ha,Yang:2013coa}, in addition to more recent observations of HII galaxies and SN luminosity distances with the angular diameter distances of ultra-compact structure in radio quasars (QSO) -- for instance, see respectively \cite{Liu:2021fka,Tonghua:2023hdz}. Those analyses showed no statistically significant deviation of the standard CDDR, given the limitation of the observational data.

In spite of that, it is important to revisit the CDDR in light of recently available observations. Since we have a new sample of baryon acoustic oscillation (BAO) measurements provided by the Dark Energy Spectroscopic Instrument~\cite{DESI:2024mwx,DESI:2025zgx}, along with existing measurements by the Sloan Digital Sky Survey (SDSS)~\cite{Gil-Marin:2016wya,eBOSS:2020gbb,eBOSS:2020tmo,Carvalho:2015ica,Alcaniz:2016ryy,Carvalho:2017tuu,deCarvalho:2017xye,deCarvalho:2021azj} and Dark Energy Survey (DES)~\cite{DES:2024pwq}, it is possible to use them as standard rulers, thus providing us $D_{\text{A}}(z)$, along with recent SN compilations, e.g. the Pantheon+ and SH0ES sample~\cite{Brout:2022vxf,Riess:2021jrx} as the probe of $D_{\text{L}}(z)$. In~\cite{Favale:2024sdq}, the authors explored the consistency of the CDDR using this combination of datasets, reporting hints at a potential deviation for it within $\sim 2\sigma$ confidence level, which the authors ascribed to possible tensions between SNe and BAO -- or perhaps internal inconsistencies among the BAO data subsets. More recently, in~\cite{Teixeira:2025czm}, it was shown that departures of the CDDR could solve the $H_0$ tension.

Another possible explanation for this result relies on more exotic physical processes. In a recently proposed model, namely the minimally extended varying speed of light (meVSL)~\cite{Lee:2020zts}, we have $\tilde{c}(z) \propto (1+z)^{-b/4}$, which implies in a photon frequency shift that leads to a violation of the standard CDDR~\cite{Lee:2021xwh} -- for alternative formulations of variable speed of light models in the literature, see \cite{Gupta:2023mgg,Nguyen:2024ddb}. We choose this VSL formulation because it maintains the CMB temperature evolution law $T \propto a^{-1}$, which is well constrained by observational data~\cite{Bengaly:2020vly,Ruchika:2025sbb}. Therefore, our goal in this work is to test the validity of the CDDR by probing $\eta(z)$ as a function of the redshift, where we adopt a combination of the aforementioned SN and BAO distance measurements to constrain the meVSL parameter $b$.

The paper is organised as follows: in section 2 we describe theoretical framework, in section 3 we present the observational data. In section 4 we detail the methodology deployed in our analysis, in section 5 we show our observational results, and in section 6 we provide our discussion and concluding remarks.

\section{Theoretical Framework}\label{sec:theory}
In this section we briefly review the main quantities of the above mentioned meVSL model by following~\cite{Lee:2021xwh}. The approach consists of assuming a flat Universe described by the Friedmann-Lemaître-Robertson-Walker (FLRW) metric and the reduced Hubble parameter can be written as
\begin{equation}
    E^{2}(a) = \left[\frac{H(a)}{H_{0}}\right]^{2} = (\Omega_{m0}a^{-3}+\Omega_{X0}a^{-3(1+\omega)a^{\frac{b}{2}}}) \equiv E^{\text{(GR)}^{2}}a^{\frac{b}{2}} \, ,
\end{equation}
where a subscript $0$ on each parameter denotes its value at the present epoch, $H_{0}$ is the present value of the Hubble parameter, i.e., the Hubble Constant, $\Omega_{i0} \equiv \rho_{i0}/ \rho_{cr0}$ is the present mass density contrast of the $i$-component, $\omega$ is the dark energy equation of state, and $b$ is a model parameter of meVSL which changes the speed of light as
\begin{equation}
    \label{eq:czmodelo}
    \tilde{c}(z) = \tilde{c}_{0}(1+z)^{-\frac{b}{4}} \; ,
\end{equation}
where $\tilde{c}_{0}$ represents the current value of the speed of light, as measured in Earth laboratories. Eq. (\ref{eq:czmodelo}) describes the joint variations of all related physical constants to satisfy the Lorentz invariance, thermodynamics, and Bianchi identity.

The transverse comoving distance $D_{\text{M}}$ is given by

\begin{equation}
D_{\text{M}}(z) = D_{\text{M}}^{\text{(GR)}}(z) = D_{\text{C}} \equiv \frac{\tilde{c}_{0}}{H_{0}}\int_{0}^{z}\frac{dz'}{\text{E}^{\text{(GR)}}(z')} = D_{C}^{\text{GR}}(z),
\; ,
\end{equation}
where the transverse comoving distance in the meVSL is the same as in GR. Consequently, the comoving distance $D_{\text{C}}$ in the meVSL is also equal to $D_{\text{M}}(z)$, since we are assuming a flat universe.

Moreover, still by following~\cite{Lee:2021xwh}, we can obtain the relation between the luminosity distance and the transverse comoving distance as 
\begin{equation}
    D_{\text{L}}(z) = (1+z)^{1-\frac{b}{8}} D_{\text{M}}(z) = (1+z)^{2-\frac{b}{8}} D_{\text{A}}(z),
\end{equation}
where $D_{\text{M}}(z) = (1+z)D_{\text{A}}(z)$ and hence the standard CDDR is modified in the meVSL model as follows
\begin{equation}
\label{eq:etazmodelo}
    \eta(z) \equiv \frac{D_{\text{L}}}{D_{\text{A}}(1+z)^{2}} = (1+z)^{-\frac{b}{8}}.
\end{equation}

In Eq.~\eqref{eq:etazmodelo}, $b$ denotes the parameter that mediates the speed of light variation -- see Eq.~\eqref{eq:czmodelo}. So, by the same equation, any statistically significant departure of $b=0$ would imply a violation of the CDDR and an evidence for a new physics. Therefore, we will constrain this parameter $b$ through the CDDR by using cosmic observations and statistical methods that will be detailed as follows.

\section{Observational Data}\label{sec:data_sim}

The latest SN compilation, namely the Pantheon+ and SH0ES data-set \cite{Brout:2022vxf} (see also \cite{Riess:2021jrx}), provides 1701 light curve measurements of 1550 distinct SN objects in the redshift interval $0.001 < z < 2.26$. Hence, we have 1701 measurements of SN apparent magnitudes, $m_{\text{B}}$, which can be combined with the determination of the SN absolute magnitude, $M_{\text B}$, in order to compute their luminosity distances according to
\begin{equation}\label{eq:DLz}
    D_{\text{L}}(z)=10^{\frac{m_{\text{B}}(z)-M_{\text{B}}-25}{5}} \,.
\end{equation}
where 
\begin{equation}\label{eq:MB}
    M_{\text{B}} = -19.25 \pm 0.03 \,,
\end{equation}
as reported in~\cite{SCOLNIC}. Note that we will neglect the full covariance matrix of the SN apparent magnitude uncertainties for computational reasons, however, we verified that this choice does not impact our final results.

\begin{table}[!t]
\centering
\caption{The 3D BAO data points used in this work.}
\vspace{0.3cm}
\begin{tabular}{|c|c|c|c|}
\hline 
Survey & $z$ & $D_{\text{A}} /r_{\text{d}}$ & References \\ 
\hline 
BOSS DR12 & $0.32$ & $6.5986 \pm 0.1337$ & \cite{Gil-Marin:2016wya} \\ 
BOSS DR12 & $0.57$ & $9.389 \pm 0.103$ & \cite{Gil-Marin:2016wya} \\
eBOSS DR16Q & $1.48$ & $12.18 \pm 0.32$ & \cite{eBOSS:2020gbb} \\
Ly$\alpha$-F eBOSS DR16 & $2.334$ & $11.25^{+0.36}_{-0.33}$ & \cite{eBOSS:2020tmo} \\ 
\hline
LRG1 DESI Y2 & $0.51$ & $8.998 \pm 0.1375$ & \cite{DESI:2025zgx} \\ 
LRG2 DESI Y2 & $0.71$ & $10.144 \pm 0.1376$ & \cite{DESI:2025zgx} \\ 
LRG3+ELG1 DESI Y2 & $0.93$ & $11.178 \pm 0.1101$ & \cite{DESI:2025zgx} \\
ELG2 DESI Y2 & $1.32$ & $11.898 \pm 0.2100$ & \cite{DESI:2025zgx} \\
QSO DESI Y2 & $1.48$ & $12.306 \pm 0.4813$ & \cite{DESI:2025zgx} \\
Ly$\alpha$-F DESI Y2 & $2.33$ & $11.708 \pm 0.2909$ & \cite{DESI:2025zgx} \\
\hline
\end{tabular} 
\label{tab:baoanisotropic_measurements}
\end{table}

\begin{table}[!t]
\centering
\caption{The 2D BAO data points used in this work.}
\vspace{0.3cm}
\begin{tabular}{|c|c|c|c|}
\hline 
Survey & $z$ & $\theta_{\text{BAO}}$ [deg] & References \\ 
\hline 
SDSS DR12 & $0.110$ & $19.8 \pm 3.26$ & \cite{deCarvalho:2021azj} \\ 
SDSS DR7 & $0.235$ & $9.06 \pm 0.23$ & \cite{Alcaniz:2016ryy} \\ 
SDSS DR7 & $0.365$ & $6.33 \pm 0.22$ & \cite{Alcaniz:2016ryy} \\
SDSS DR10 & $0.450$ & $4.77 \pm 0.17$ & \cite{Carvalho:2015ica} \\
SDSS DR10 & $0.470$ & $5.02 \pm 0.25$ & \cite{Carvalho:2015ica} \\ 
SDSS DR10 & $0.490$ & $4.99 \pm 0.21$ & \cite{Carvalho:2015ica} \\ 
SDSS DR10 & $0.510$ & $4.81 \pm 0.17$ & \cite{Carvalho:2015ica} \\ 
SDSS DR10 & $0.530$ & $4.29 \pm 0.30$ & \cite{Carvalho:2015ica} \\
SDSS DR10 & $0.550$ & $4.25 \pm 0.25$ & \cite{Carvalho:2015ica} \\
SDSS DR11 & $0.570$ & $4.59 \pm 0.36$ & \cite{Carvalho:2017tuu} \\
SDSS DR11 & $0.590$ & $4.39 \pm 0.33$ & \cite{Carvalho:2017tuu} \\
SDSS DR11 & $0.610$ & $3.85 \pm 0.31$ & \cite{Carvalho:2017tuu} \\
SDSS DR11 & $0.630$ & $3.90 \pm 0.43$ & \cite{Carvalho:2017tuu} \\
SDSS DR11 & $0.650$ & $3.55 \pm 0.16$ & \cite{Carvalho:2017tuu} \\
DES Y6 & $0.850$ & $2.932 \pm 0.068$ & \cite{DES:2024pwq} \\
BOSS DR12Q & $2.225$ & $1.77 \pm 0.31$ & \cite{deCarvalho:2017xye} \\
\hline
\end{tabular} 
\label{tab:baoangular_measurements}
\end{table}

As our probe of the angular diameter distance, we utilise measurements of the Baryon Acoustic Oscillations (BAO) obtained from the large-scale clustering of cosmic objects and measured by many different surveys, as presented below. Similarly to~\cite{Favale:2024sdq}, we adopt two distinct BAO datasets, which consist of measurements of the anisotropic (three-dimensional) and transverse (two-dimensional) BAO signal, which we will hereafter refer to as 3D BAO and 2D BAO, respectively, as listed in Tables~\ref{tab:baoanisotropic_measurements} (3D BAO) and~\ref{tab:baoangular_measurements} (2D BAO). Note that the 3D BAO subsample by DESI was updated to its most recent release, named DESI Y2~\cite{DESI:2025zgx}.


\section{Methodology}

Since we can use the comoving sound horizon at the drag epoch $r_{\text{d}}$ as a cosmic standard ruler~\cite{Aghanim}, galaxy surveys have been able to measure its angular scale as
\begin{equation}
    \label{eq:theta}
    \theta(z) = \frac{r_{\text{d}}}{D_{\text{M}}(z)}
\end{equation}
at various redshifts. Because the transverse (2D) BAO data listed in Table \ref{tab:baoangular_measurements} are given in terms of $\theta_{\text{BAO}}$, we can convert it to angular diameter distance measurements using the following relation
\begin{equation}
    \label{eq:DAz}
    D_{\text{A}}(z) = \frac{r_{\text{d}}}{(1+z)\theta_{\text{BAO}}}
\end{equation}
and we can obtain their respective uncertainties through standard error propagation, given as follows,
\begin{equation}
    \label{eq:sigmaDA_2D}
    \sigma^{2}_{D_{\text{A}}(z)} = \left[\frac{r_{\text{d}}}{\theta^{2}_{\text{BAO}}(z)(1+z)}\right]^2 \sigma^{2}_{\theta_{\text{BAO}}(z)} + \left[\frac{1}{\theta_{\text{BAO}}(z)(1+z)}\right]^2 \sigma^{2}_{r_{\text{d}}}.
\end{equation}
As for the anisotropic (3D) BAO, we can obtain $D_{\text{A}}(z)$ by simply multiplying $D_{\text{A}}(z)/r_\text{d}$ by the value of $r_{\text{d}}$, so that their respective uncertainties read
\begin{equation}
    \label{eq:sigmaDA_3D}
    \sigma^{2}_{D_{\text{A}}(z)} = \left[r_{\text{d}}\right]^2 \sigma^2_{(D_{\text{A}}(z)/r_{\text{d}})} + \left[\frac{D_{\text{A}}(z)}{r_{\text{d}}} \right]^2 \sigma^2_{r_{\text{d}}}.
\end{equation}
The sound horizon value adopted in this work, unless stated otherwise, consists on the best-fitted values from low-$z$ cosmological probes (SNe and BAO) when assuming a $H_0$ prior from SH0ES (as of 2018 values and data), i.e., $r_{\text{d}} = 136.4 \pm 3.5 \, \text{Mpc}$ -- see section 5.4 in~\cite{Aghanim}. We assume this value, rather than the Planck CMB's best fit ($r_{\text{d}} = 147.05 \pm 0.3 \, \text{Mpc}$), because the latter is not consistent with the implicit $H_0$ assumption we make when using the SN absolute magnitude value, as in Eq.~\eqref{eq:MB}, provided by Pantheon+ and SH0ES, which is $\sim 5\sigma$ confidence level away from the $H_0$ CMB measurement. This would lead to a spurious departure of the CDDR -- see~\cite{Keil:2025ysb} for a recent discussion about this topic. In addition, the Hubble Constant measurement by SH0ES only assumes cosmography, whereas Planck's Hubble Constant (and consequently the sound horizon) measurements need to assume $\Lambda$CDM, or other dark energy models. This justifies our choice for the sound horizon as $r_{\text{d}} = 136.4 \pm 3.5 \, \text{Mpc}$.

In order to probe the CDDR, we again recall the $\eta(z)$ function, defined as
\begin{equation}
\label{eq:etaz2modelo}
    \eta(z) = \frac{D_{\text{L}}(z)}{(1+z)^{2}D_{\text{A}}(z)},
\end{equation}
whose error propagation yields
\begin{equation}
\label{eq:sigma_etaz}
    \sigma^{2}_{\eta(z)} = \left[\frac{1}{(1+z)^{2}D_{\text{A}}(z)}\right]^2\sigma^{2}_{D_{\text{L}}(z)}+ \left[\frac{D_{\text{L}}(z)}{(1+z)^{2}D^{2}_{\text{A}}(z)}\right]^2\sigma^{2}_{D_{\text{A}}(z)},
\end{equation}
where the angular diameter distance uncertainties correspond to Eqs.~\eqref{eq:sigmaDA_2D} and~\eqref{eq:sigmaDA_3D} for the 2D and 3D BAO cases, respectively, whereas the luminosity distance uncertainty is given by
\begin{equation}
\label{eq:sigma_DL}
\sigma_{D_{\text L}(z)} = \frac{\log{(10)}}{5}10^{\frac{m_{\text{B}}(z)-M_{\text{B}}-25}{5}}\sigma_{m_{\text{B}}(z)}.
\end{equation}

Nonetheless, in order to obtain $\eta(z)$, we need SN and BAO distance measurements in the exact same redshift, which is usually not available. We circumvent this problem by employing a Gaussian Process (GP) reconstruction using the well-known {\sc GaPP} code~\cite{MSeikel} on the SN luminosity distance measurements $D_{\text L}(z)$. Thus, we can match $D_{\text{L}}$ with $D_{\text{A}}$ at the same redshift of the latter in a model-independent way, i.e., without prior assumptions on a cosmological model which could bias our analysis. Note also that the GP hyperpameters were optimised in this analysis, and that we set no specific priors on these quantities. 

Then we can perform an interpolation across the SN data points, i.e., 2400 bins in the redshift range $0.1 < z < 2.5$, using the Squared Exponential kernel, to match their redshifts with those of the BAO measurements -- and most importantly, without any prior assumption on the underlying cosmological model -- to finally obtain $\eta(z)$. 

The final step of our method is to estimate the meVSL parameter $b$. To accomplish this, we perform a $\chi^2$ minimisation according to
\begin{equation}
\label{eq:chi2_b}
\chi^2 = \sum_{i}\frac{\left[\eta_{i}(z) - (1+z_i)^{-b/8}\right]^2}{\sigma^2_{\eta_{i}(z)}},
\end{equation}
for $b$ assuming a flat prior at $-3 \leq b \leq 3$, and the index $i$ denoting each $\eta(z)$ inference, i.e., number of BAO data points in the corresponding sample. This prior is wide enough to encompass at least the $3\sigma$ confidence regions for any dataset combination, which justifies its choice. It is worth noticing that, after the redshift matching procedure through GP previously described, we split the data into five different sub-samples, specifically (i) a sub-sample with the combination of the 2D+3D BAO (DESI) data, (ii) a sub-sample with the 2D+3D BAO (SDSS) data, (iii) sub-sample with the 2D BAO data only, as well as (iv) the 3D BAO (DESI) and (v) 3D BAO (SDSS) ones, by the same fashion as in~\cite{Favale:2024sdq}.

In addition, we perform another $\chi^2$ test (for all those BAO subsamples) relative to the standard CDDR relation by means of
\begin{equation}
\label{eq:chi2_eta0}
\chi^2 = \sum_{i}\frac{\left[\eta_{}(z) - \eta_0\right]^2}{\sigma^2_{\eta_{i}(z)}} \; .
\end{equation}
In this second approach we assume a constant value for the CDDR parameterisation ($\eta_0$)~\cite{Uzan2004, DeBernardis:2006ii}. There is a two-fold motivation for this approach, first it is the most simple case for a mathematical function to describe such deviation, and second it is not redshift dependent, so we can contrast with the previous parameterisation.

After both analyses we also perform a comparison between these cases.

\section{Results}
\begin{figure}[!t]
\includegraphics[scale=0.40]{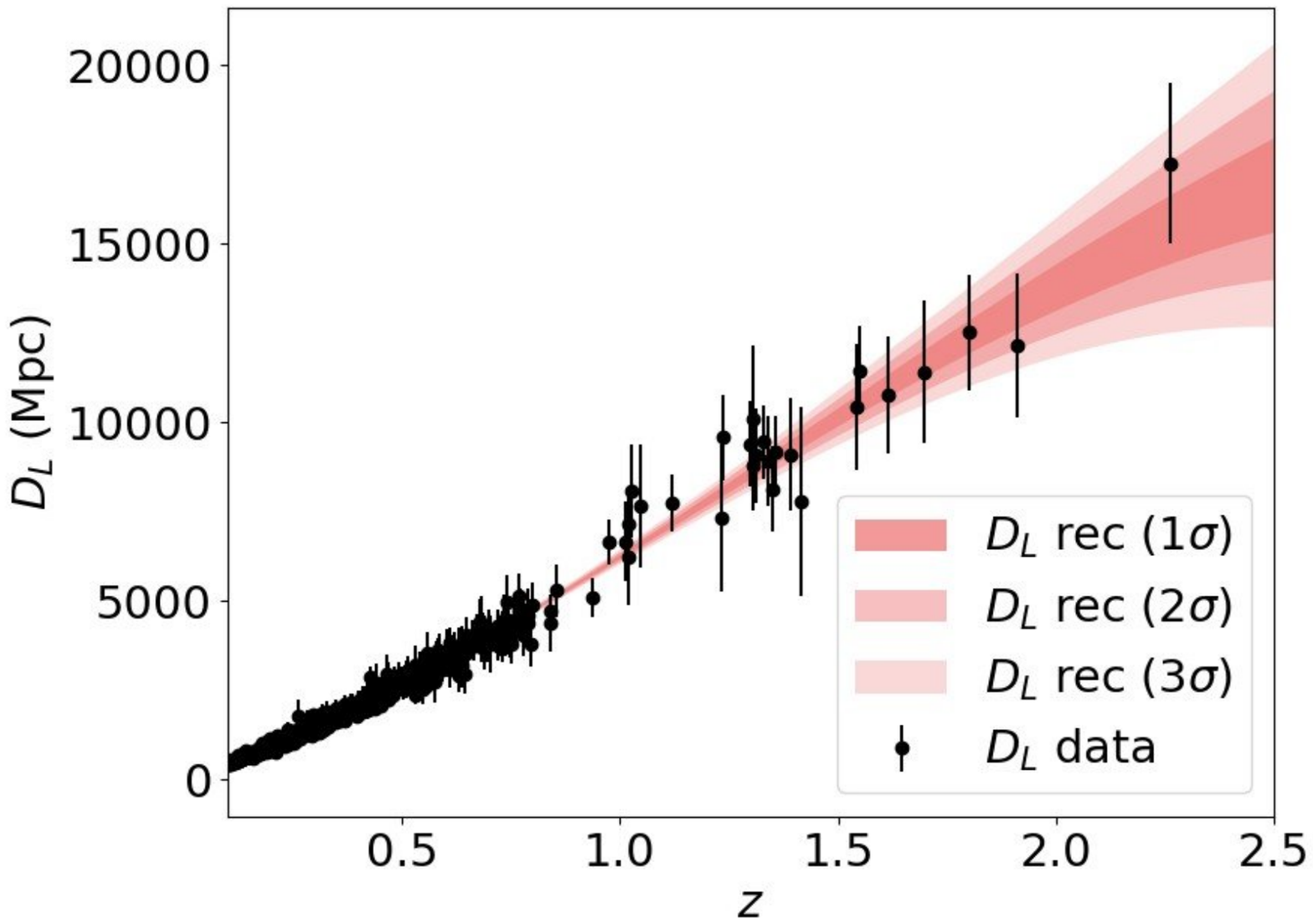}
\caption{Gaussian Process reconstruction of the luminosity distance, $D_{\text{L}}(z)$, as obtained by Eqs. \ref{eq:DLz}, \ref{eq:MB} and \ref{eq:sigma_DL}, using the Squared Exponential kernel. The black dots with error bars in the $D_{\text{L}}(z)$ plot represent the observational data, and the pink curves denote the $1$, $2$, and $3\sigma$ confidence level of the reconstructions, respectively from the darker to the
lighter shade.}
\label{fig:dlrec}
\end{figure}

\begin{figure}[!t]
\includegraphics[scale=0.40]{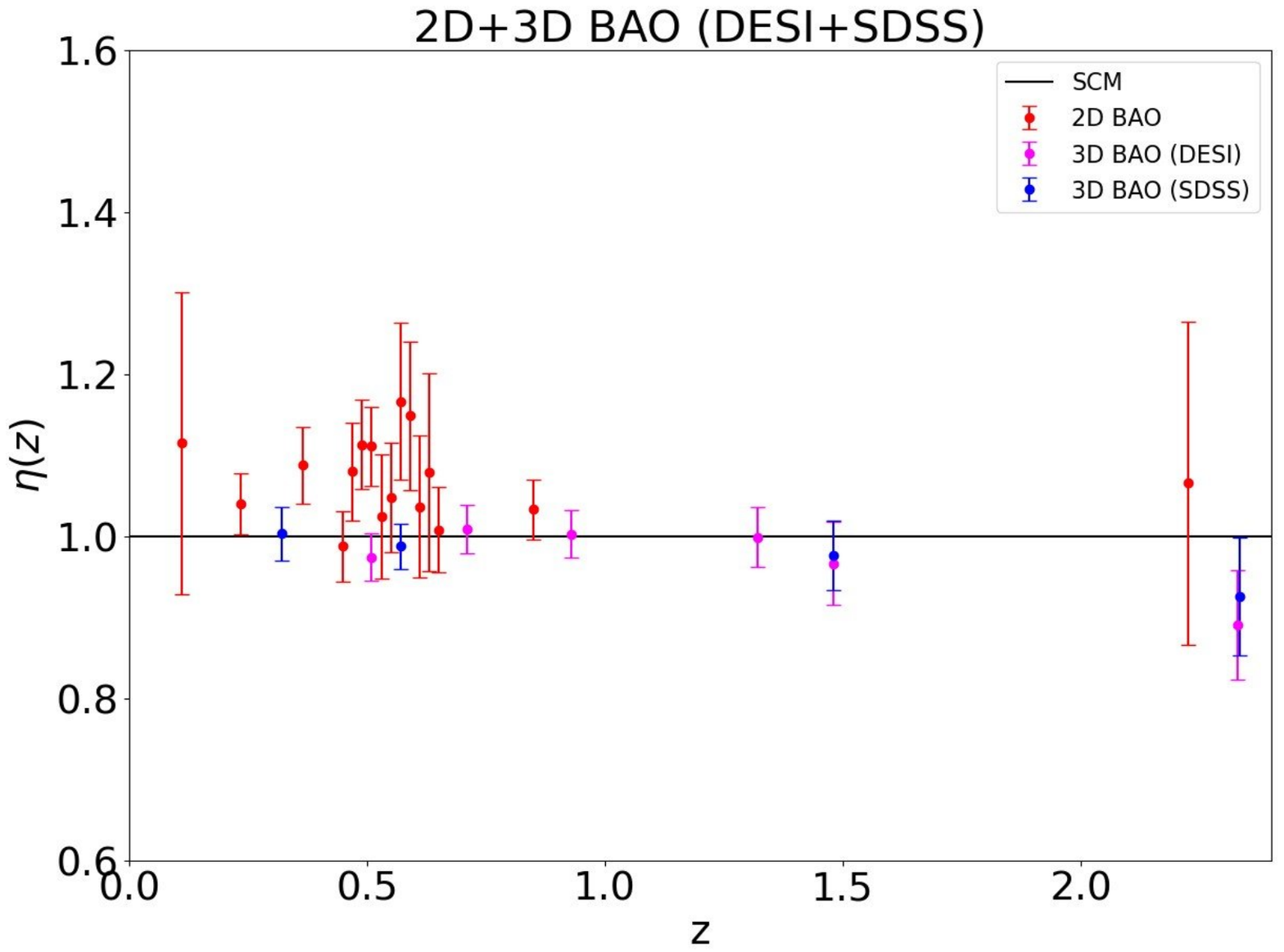}
\caption{$\eta(z)$ measurements at $2\sigma$ confidence level obtained from the Pantheon+ and SH0ES SNe with the joint BAO data (2D, 3D BAO (DESI) and 3D BAO (SDSS). The blue, red and magenta data points represent the SNe data combined with the 3D BAO (SDSS), SNe with 2D BAO and SNe with 3D BAO (DESI), respectively. The horizontal line represents the standard CDDR, i.e., $\eta(z) = 1$.}
\label{fig:eta_blik_2d3dbao}
\end{figure}

In the Figure \ref{fig:dlrec} we show the Gaussian Process reconstruction of the luminosity distance, $D_{\text{L}}(z)$, as obtained by Eqs. \ref{eq:DLz}, \ref{eq:MB} and \ref{eq:sigma_DL}, using the Squared Exponential kernel. The $\eta(z)$ distribution of all points, as a function of the redshift, can be found in Figure~\ref{fig:eta_blik_2d3dbao}. In the panel we show the whole distribution, where the blue points correspond to the combination of SNe and 3D BAO (SDSS), and the red points correspond to the combination of SNe and 2D BAO and the magenta points correspond to the combination of SNe with 3D BAO (DESI).

\begin{table}[!h]
\centering
\caption{Best fit results for $b$, as in $\eta(z) = (1+z)^{-b/8}$. The first column displays the combination of data-sets, the second column provides the best-fitted values for $b$ and their uncertainties in $1\sigma$, the third column stands for the reduced $\chi^2$, named $\chi^2_\nu$, and the fourth column gives their deviation from $b=0$.}
\vspace{0.3cm}
\begin{tabular}{|c|c|c|c|c|}
\hline 
data-sets (SNe+) & $b \pm \sigma_{b}$ & $\chi^2_{\nu}$ & dev ($b = 0$) \\ 
\hline 
2D+3D BAO (DESI) & $-0.147 \pm 0.142$ & $1.274$ & $1.04\sigma$  \\ 
2D+3D BAO (SDSS) & $-0.281 \pm 0.177$ & $1.208$ & $1.59\sigma$ \\ 
2D BAO & $-0.911 \pm 0.255$ & $0.776$ & $3.57\sigma$ \\ 
3D BAO (DESI) & $0.173 \pm 0.174$ & $0.587$ & $0.99\sigma$ \\ 
3D BAO (SDSS) & $0.268 \pm 0.256$ & $0.134$ & $1.05\sigma$ \\ 
\hline
\end{tabular} 
\label{tab:b_bestfits}
\end{table}


\begin{table}[!h]
\centering
\vspace{0.3cm}
\caption{Best fit results for $\eta_0$, as in $\eta(z) = \eta_0 = \mathrm{constant}$. The first column displays the combination of data-sets, the second column provides the best-fitted values for $\eta_0$ and their uncertainties in $1\sigma$, the third column stands for the reduced $\chi^2$, named $\chi^2_\nu$, and the fourth column gives their deviation from $\eta_0=1$.}
\begin{tabular}{|c|c|c|c|c|}
\hline 
data-sets (SNe+) & $\eta_0 \pm \sigma_{\eta_0}$ & $\chi^2_{\nu}$ & dev ($\eta_0 = 1$) \\ 
\hline 
2D+3D BAO (DESI) & $1.021 \pm 0.010$ & $1.118$ & $2.09\sigma$  \\ 
2D+3D BAO (SDSS) & $1.029 \pm 0.011$ & $1.005$ & $2.64\sigma$ \\ 
2D BAO & $1.055 \pm 0.014$ & $0.637$ & $3.93\sigma$ \\ 
3D BAO (DESI) & $0.988 \pm 0.014$ & $0.658$ & $0.86\sigma$ \\ 
3D BAO (SDSS) & $0.987 \pm 0.018$ & $0.339$ & $0.72\sigma$ \\ 
\hline
\end{tabular} 
\label{tab:eta0_bestfit_lowz}
\end{table}


\begin{figure}[!t]
\includegraphics[scale=0.40]{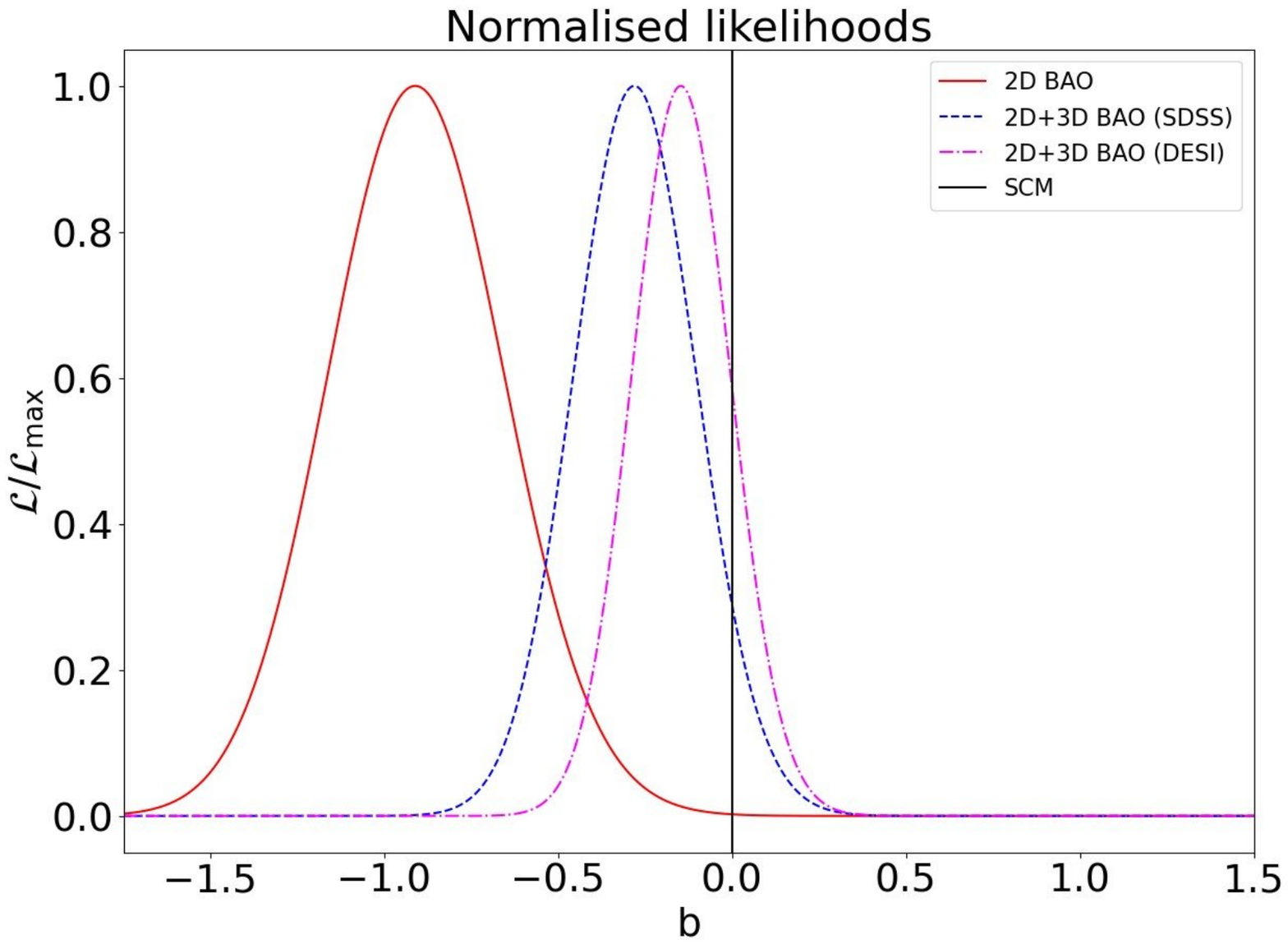}
\includegraphics[scale=0.40]{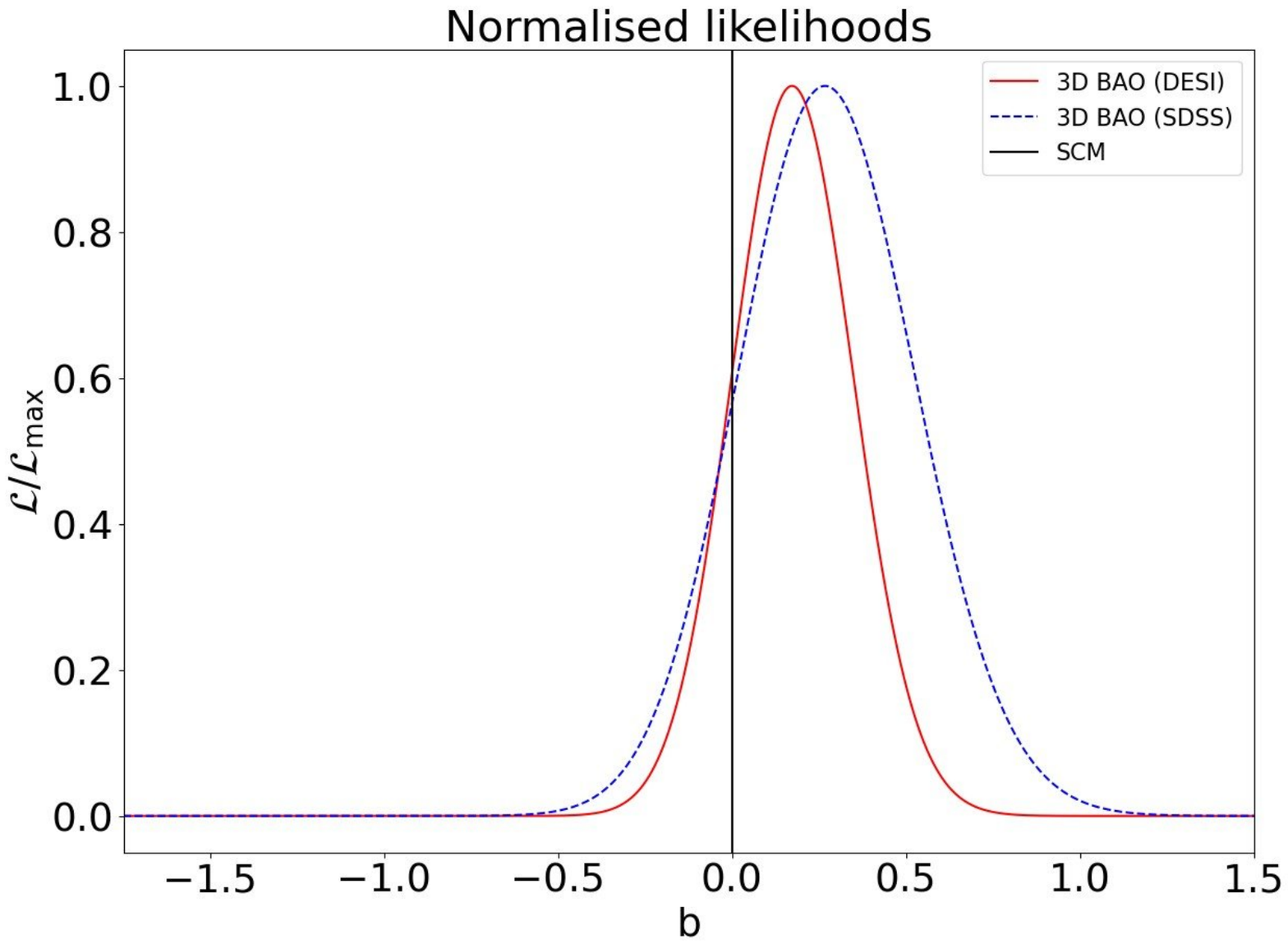}
\caption{The normalised likelihoods for the $b$ parameter assuming SNe along with ({\it upper panel:}) 2D+3D BAO (SDSS) joint sample (blue dashed line), 2D BAO (red continuous line), and 2D+3D BAO (DESI) joint sample (magenta dashed-dotted line), ({\it lower panel:}) 3D BAO (SDSS, blue dashed line) and 3D BAO (DESI, red continuous line). The vertical line denotes the standard CDDR case, i.e., $b=0$.}
\label{fig:eta_blik_3dbao}
\end{figure}

\begin{figure}[!t]
\includegraphics[scale=0.40]{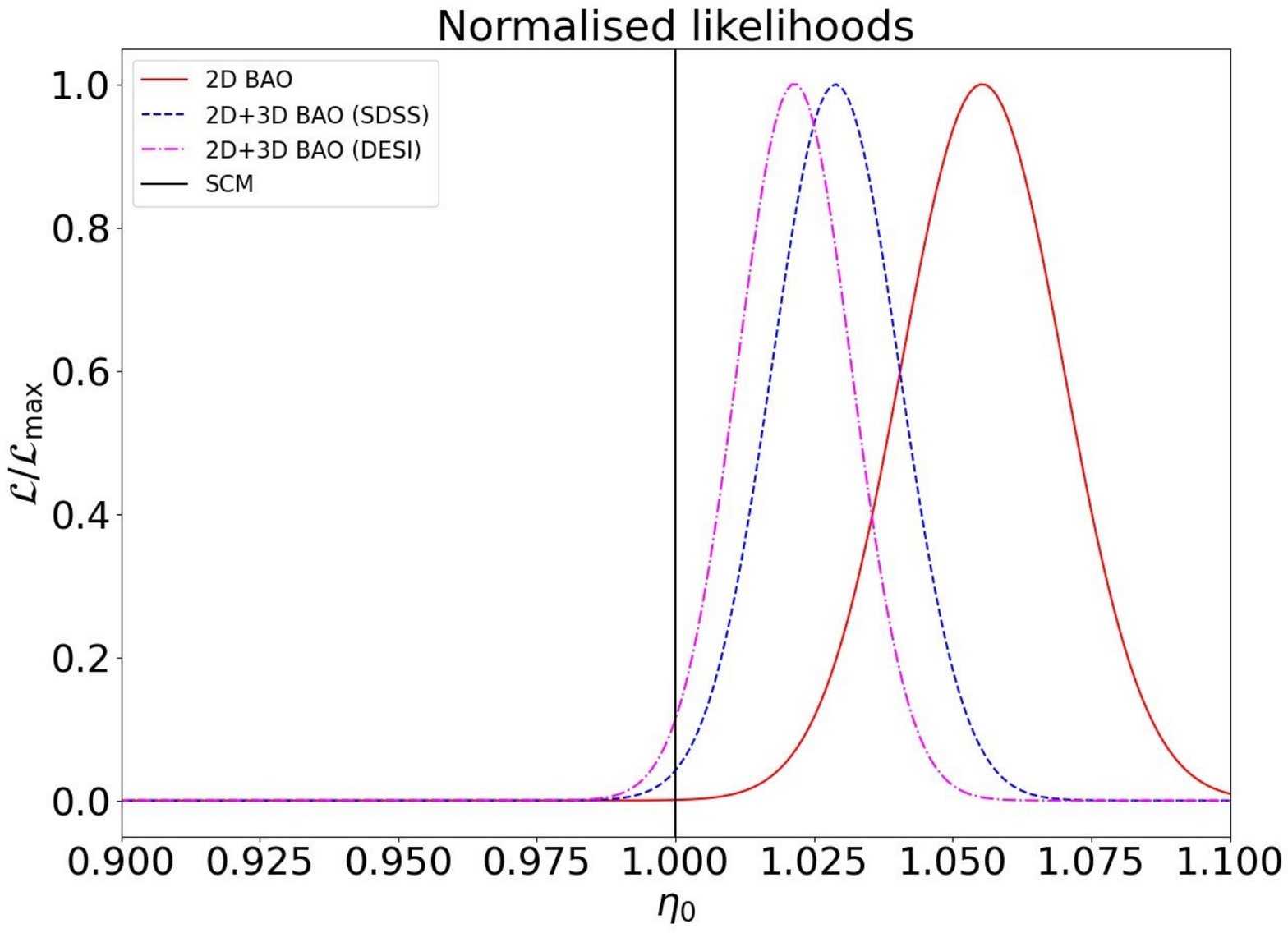}
\includegraphics[scale=0.40]{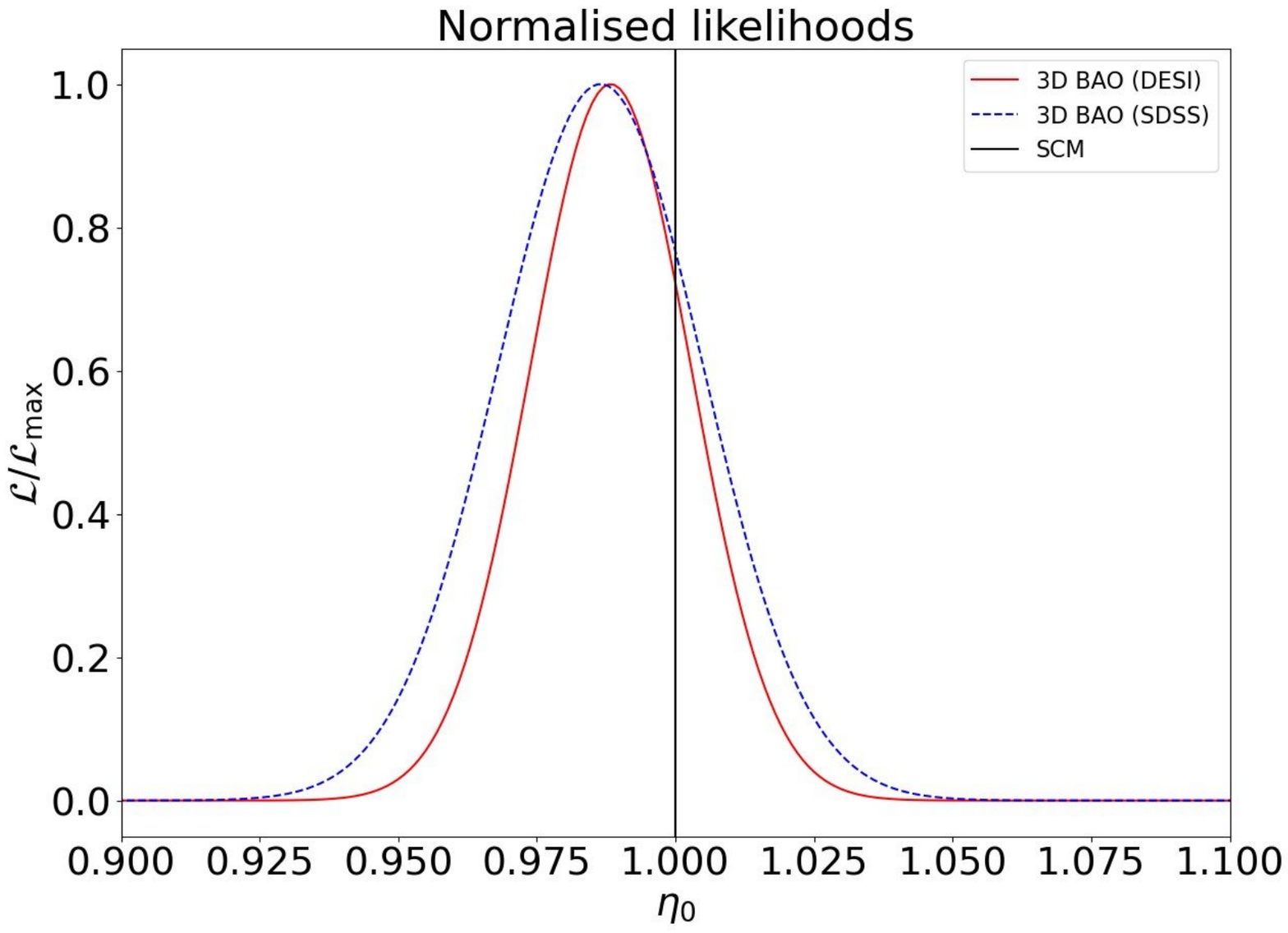}
\caption{The normalised likelihoods for the $\eta_0$ parameter assuming SNe along with ({\it upper panel:}) 2D+3D BAO (SDSS) joint sample (blue dashed line), 2D BAO (red continuous line), and 2D+3D BAO (DESI) joint sample (magenta dashed-dotted line), ({\it lower panel:}) 3D BAO (SDSS, blue dashed line) and 3D BAO (DESI, red continuous line). The vertical line denotes the standard CDDR case, i.e., $\eta_0 = 1.0$}
\label{fig:eta_0lik_3dbao}
\end{figure}

In Table~\ref{tab:b_bestfits} we show the best fits and $1\sigma$ (second and third columns, respectively) deviation we found for the meVSL parameter $b$ with respect to each sub-sample of BAO used (first column).  As we can see in the third column the reduced $\chi^2$ rely on an acceptable range, close to the unity. In the fourth column we can find the deviation of the best fit relative to the scenario where the CDDR is valid (i.e. $b = 0$). In almost all cases, the deviation is around $1\sigma$, making the results completely compatible with the validity of the CDDR. The most discrepant deviation can be found for the sub-sample comprised only with the 2D data. This result is compatible with recent works as~\cite{Favale:2024sdq} that also shows some particular features when the angular BAO is used. In order to better address this question, further investigations will be performed in future works.

As mentioned in the previous section, we also performed the analysis by assuming a parameterisation of $\eta(z) = \eta_0$ and the results can be found in Table~\ref{tab:eta0_bestfit_lowz}. We show the best fits and $1\sigma$ deviation with respect to each sub-sample of BAO used. We also can find the deviation of the best fit relative to the scenario where the CDDR is valid (i.e. $\eta_0 = 1$). Although the combinations of 2D + 3D BAO (DESI) and 2D + 3D BAO (SDSS) data are slightly deviating from the validity of the CDDR, the most discrepant deviation can also be found for the sub-sample comprised only with the 2D data. The correspond plots for each one of the previous analyses can be found in Figures~\ref{fig:eta_blik_3dbao} and~\ref{fig:eta_0lik_3dbao}.

\section{Conclusions}

The cosmic distance duality relation (CDDR) is one of the fundamental relations in cosmology which plays a crucial role in astronomical observations. Any deviation from it indicates that either the spacetime is not described by a metric theory of gravity or new physics beyond the one we understand. Given the availability of new improved observational data with better precision, it is important to test this relation. Supernova and baryon acoustic oscillation measurements can be considered effective observational data for testing CDDR, as they can provide precise measurements of the luminosity distance, $D_{\text{L}}(z)$, and angular diameter distance, $D_{\text{A}}(z)$, respectively. 

In this work, we probe the cosmic distance duality relation (CDDR) by comparing angular diameter distance measurements, provided by the transverse (two-dimensional, 2D) and anisotropic (three-dimensional, 3D) baryon acoustic oscillations (BAO) data from SDSS and DESI, with luminosity distance measurements from the Pantheon+ and SH0ES type Ia Supernova (SNe) compilation. The Gaussian Process reconstruction is employed on the SN data to match the $D_{\text{L}}(z)$ from SNe with the redshifts of the $D_{\text{A}}(z)$ from BAO, so that we can compute $\eta(z)$ at each redshift where there is both data for SNe and BAO. 

The function $\eta(z) \equiv D_{\text{L}}/D_{\text{A}}(z)(1+z)^{-2} = 1$ is used to verify the possible deviation at any redshift, so we tested it using two approaches: on one hand we verify the the so-called minimally extended varying speed of light (meVSL) model, where the above relation is modified as $\eta(z) = (1+z)^{-b/8}$ and, on the other hand, we assume a simple constant function where $\eta(z) =\eta_0$, where the case $\eta_0=1$ recovers the standard CDDR.

In the first approach, we estimate the best-fit value for $b$ through the maximum likelihood estimator, where we find no deviation of the standard CDDR relation, within 1$\sigma$ confidence level, when considering SNe combined with any 3D BAO subsamples, either by DESI or SDSS, as well as considering SNe combined with the joint 2D and 3D BAO datasets. However, we find a $> 3\sigma$ confidence level deviation from the standard CDDR, in favour of a varying speed of light as in the meVSL framework, when these SNe are combined with the 2D BAO dataset.

In the second approach, we also verify the results for $\eta_0$, and we found results with similar conclusions compared with the previous one. It means that while the 3D subsamples and the combination of 2D and 3D subsamples lead to a maximum of $2.6\sigma$ deviation, this deviation increases to $\sim 4 \sigma$ when the 2D BAO data only is considered. Such results might be ascribed to unaccounted systematics in the 2D BAO, as well as internal inconsistencies between the 2D and 3D BAO datasets, e.g. the preference for different $r_{\rm d}$ values, as suggested in~\cite{Carvalho:2015ica}. Moreover, it is worth mentioning that these results are valid for a specific choice of priors on the BAO sound horizon scale $r_{\rm d}$, as well as the SN absolute magnitude $M_{\rm B}$, which in turn is related to the Hubble Constant $H_0$ by SH0ES. A more thorough assessment on how those parameters can affect the meVSL parameter $b$, especially in light of the $H_0$ tension and a possible discrepancy between the 2D and 3D BAO measurements, will be pursued in a future work. 

Therefore, the significance of the evidence for the meVSL model in the case of the 2D BAO with SNe is still under debate, and further investigations will be performed in order to clarify whether this is due to degeneracy of the distance calibrator priors, inner disagreement between the available datasets, or a hint at novel physics beyond the standard model.

{\it Acknowledgments:} JS acknowleges financial support from Coordenação de Aperfeiçoamento de Pessoal de Nível Superior (CAPES). CB acknowledges financial support from Funda\c{c}\~ao \`a Pesquisa do Estado do Rio de Janeiro (FAPERJ). RSG thanks financial support from FAPERJ grant No. 260003/005977/2024 - APQ1. 
\newpage


\end{document}